"Embarrassingly Causal": Causal Use of Associational Data in Magic The Gathering Drafts

Mark Louie F. Ramos, Ph.D.
*Department of Health Policy and Administration, The Pennsylvania State University, University Park, PA*
*mlr6219@psu.edu*

Abstract

Observational data are often used to answer causal questions, yet the legitimacy of doing so is often argued to hinge on strong, domain-supported assumptions about underlying causal structure with limited guidance on how much domain-knowledge support should exist to justify including a causal edge of interest in a directed acyclic graph (DAG). We introduce the criterion of "embarrassingly causal" scenarios, where the existence of an exposure → outcome relationship is so uncontroversial that the assumptions needed to include the corresponding causal edge in a DAG can be reasonably made. Using the case of Magic: The Gathering booster draft decisions and gameplay outcomes, we show how purely observational data from 17Lands.com are widely and effectively used to guide draft choices despite substantial confounding, selection effects, and post-treatment conditioning. We argue that the embarrassingly causal quality is a sufficient condition for justifying the construction of causal estimands and the collection of observational data to estimate them. Correspondingly, we provide guidance on evaluating observational causal inference assumptions for authors, reviewers, and readers.

1. Introduction

There is ongoing debate over the extent to which causal structure can be learned from observational data. In machine learning, a growing literature on causal discovery argues that, under sufficiently strong structural assumptions such as faithfulness, causal sufficiency, or independence of causal mechanisms, statistical patterns in observational data may partially identify causal graphs or even latent causal variables [1–3] Within this framework, causal structure is treated as an object that may be discoverable, at least in part, from appropriately modeled data. In contrast, much of contemporary causal inference in statistics and epidemiology emphasizes that causal directed acyclic graphs (DAGs) primarily encode scientific assumptions about the data-generating process (dgp) and cannot themselves be justified by observational data alone [4–6]. In this perspective, researchers use substantive knowledge about the dgp to assume a specific DAG and estimate causal quantities identified under this assumption.

Amid this debate lies persistent difficulties in communicating the foundational limits of causal inference from observational data to both the general public and even to practicing quantitative researchers who encounter or use observational causal inference tools. Recent work suggests that misunderstandings commonly arise at both extremes: some interpret "correlation is not causation" as an absolute prohibition [7, 8], missing the nuance that correlations can reflect causal relationships under assumptions, while others overread associations as causal without justification [9] or assume that turning an association into a causal effect is a purely statistical or mathematical exercise [10].

It is in this context that this paper is situated. We take the position that causal inference from observational data necessarily relies on strong, explicit a priori assumptions about causal structure that must be articulated prior to decisions about acquiring or analyzing data and should heavily inform such decisions. In this sense, causal assumptions are not merely considerations that are applied post hoc, like the Bradford-Hill criteria for when observed associational evidence may be treated as causal [11], but foundational commitments that determine whether an observational dataset may be relevant to a causal question in the first place. As such, a central methodological concern is being able to qualify the reasonableness of these assumptions. Many sources point to the importance of domain knowledge from subject-matter experts in justifying a causal edge in a DAG but there is no guidance on how much domain knowledge is sufficient [5, 12]. In fact, it is somewhat meaningless to think about this in terms of how many subject-matter experts one can get to agree that it is reasonable to include a specific causal edge in a graph or make a potential outcomes assumption, and yet there is somewhat a kernel of truth to that since if practically everyone in the domain agrees that a specific causal edge is obviously reasonable to include, then that assumption is defensible based on existing domain knowledge. Towards this end, we define the "embarrassingly causal" criterion to formalize a class of situations in which the causal structure between an exposure and an outcome is so strongly supported by domain knowledge that including the corresponding causal edge in a DAG is uncontroversial. While, even in these scenarios, identification of specific useful estimands may remain difficult or even impossible, and such estimands may still be practically negligible or even zero, the causal direction itself need not be treated as uncertain.

We demonstrate an embarrassingly causal scenario in the context of using observational data from 17lands.com when making pick decisions for the booster draft game format of the collectible card game, Magic: The Gathering. In this setting, players repeatedly face well-defined decision points, followed by downstream game outcomes that depend on those choices. Although the data collected by 17Lands are entirely observational and subject to estimand identification and analysis issues such as substantial confounding, selection effects, and post-treatment conditioning, we show that the causal structure from pick choices to subsequent win outcomes is not plausibly in doubt, making it reasonable to collect and analyze such data for the purpose of informing intervening decisions. In the succeeding sections, we expound on the context of Magic: The Gathering booster drafts and the 17Lands dataset. Then, we illustrate the causal identification problem in this context and demonstrate how purely observational data are effectively used in practice. We examine the justification for this usage under our embarrassingly causal framework and consider the extent to which this framework is sufficient and necessary for observational causal inference in general.

2. Magic: The Gathering Booster Drafts

2.1. Magic: The Gathering

Magic: The Gathering (MTG) is a long-running collectible card game created in 1993 by mathematician Richard Garfield [13]. It has since grown into a globally recognized strategy game with millions of players and a deeply studied competitive ecosystem and trading card economy [14–17]. Each game of MTG is played with decks constructed from a large pool of cards and is almost always a zero-sum game where players take turns until one of them wins. MTG is a billion dollar brand [18] with a competitive scene that offers millions of dollars in prizes annually [19].

2.2. Booster Draft Leagues

I focus this exposition on Magic: The Gathering Arena (MTGA) booster draft leagues. MTGA is the primary digital platform for playing Magic: The Gathering. In a booster draft, eight human players are placed into a virtual pod. Each player opens a 15-card booster pack, selects one card, and passes the remaining cards to the next player. This continues until all cards from the pack have been drafted. The process is repeated for a total of three packs per person, with the direction of passing alternating in each pack. After drafting, players construct a minimum 40-card deck using the cards they selected [20]. After building their decks, players enter a queue and are matched against other players to play a round. They can be matched against any player who is in the queue, not necessarily the people they drafted with [21].

### 2.3. Causal Identification Problem

Define $C_t$ as the action of picking a card at pick $t$, where $t = 1 \ldots 45$. Next, define the indicator variable $1_{win}$, which denotes that future outcome (win or loss) of a match that the player will play using the deck drafted. We are interested in the causal effect of $C_t$ on $1_{win}$. That is, how much does picking a specific card among the choices in pick $t$ affect the probability that the player will win some future match with the deck eventually drafted. Formally, for any card $c$ present at pick $t$, we are interested in the following estimands.

$$Pr\big(1_{win} \mid do(C_t = c)\big) \quad (1)$$

$$Pr\big(1_{win} \mid do(C_t = c)\big) - Pr\big(1_{win} \mid do(C_t = c')\big) \quad (2)$$

Knowing these gives the player a clear, causally justified, strategy on what card to choose. Among the cards presented to them in pick $t = 1$, an optimal strategy is some function of (1) and (2). For example, choosing the maximum (2) among the cards meeting some minimum threshold for (1). Beyond $t = 1$, the same rule may be used but with the caveat of considering cards already chosen. The influence of this caveat grows as the draft progresses, until such a time when the player decides to "lock-in" and limit their choices only among cards that are relevant to the current pool of cards already selected.

However, estimating these quantities is very challenging. Decisions for each pick are embedded in a dynamic, history-dependent process with downstream consequences and substantial outcome noise. Likewise, the outcome of winning a match given a card is picked is influenced by a cascade of factors such as the player's and their opponent's skill and strategic tendencies, the favorability of the match based on all other cards in the players' decks and the inherent variance of the game itself.

3. Embarrassingly Causal Associations in 17Lands data

### 3.1. 17Lands data

17Lands is a large, crowd-sourced dataset collected from players of Magic: The Gathering Arena who opt in to share their draft and gameplay logs [22]. For each draft, the system records the full sequence of packs and picks encountered by the player (but not by other players in the draft who have not opted into 17Lands), the evolving contents of the player's deck, and the outcomes of the matches played with that deck. This results in a rich dataset that contains millions of pick decisions and thousands of match results across a wide range of players.

From this raw dataset, 17Lands computes various card-level summary statistics that are published on their website [23]. These include both raw frequencies, such as the number of packs in which a card has been seen, and functions of frequencies, such as Average Last Seen At (ALSA), defined as "the average pick number where this card was last seen in packs" [23]. Three statistics specifically of interest are as follows:

Games in Hand Win Rate (GIHWR): The win rate of games where an instance of the card was drawn into hand, either in the opening hand or later.

Games Not Seen Win Rate (GNSWR): The win rate in games where one or more copies of the card was in the deck but not seen at any stage of the game.

Improvement In Hand (IIH): The difference between Games in Hand Win Rate and Games Not Seen Win Rate.

These statistics are of interest because they seem to be estimators of our estimands, yet it is clear that these estimators are not unbiased for our estimands. First, GIHWR is conditioned on drawing the card, and thus is closer to the estimand:

$$Pr\bigl(1_{win} \mid do(C_t = c), 1_{draw\_c} = 1\bigr) \qquad (3)$$

While (3) is also a reasonable estimand, GIHWR is still biased for (3). For example, drawing a specific card $c$ is confounded with variables such as the number of turns in the game, which also affects $Pr(1_{win})$ depending on the totality of cards in each deck (some decks are designed to win quickly while others are designed to play longer games). In addition, some cards are much more impactful when played in earlier turns than on later turns.

On the other hand, IIH only looks visually similar to (2) but is entirely different. The estimand (2) compares the probability of winning when picking $c$ at pick $t$ compared to not doing so, but IIH conditions on having picked $c$ and compares the win rate when it is drawn compared to when it is not. Thus, the corresponding estimand for IIH is closer to

$$Pr\bigl(1_{win} \mid do(C_t = c), 1_{draw\_c} = 1\bigr) - Pr\bigl(1_{win} \mid do(C_t = c), 1_{draw\_c} = 0\bigr) \qquad (4)$$

And IIH is still biased for (4) similar to how GIHWR is biased for (3).

### 3.2. Wide and effective use of naïve associations

It is clear that both GIHWR and IIH are naïve associational quantities that can be heavily biased estimators of useful estimands (3) and (4). Earlier, it was pointed out that if good estimates of (1) and (2) were available, then the player can use simple strategies involving these to make optimal decisions for each pick. Similar arguments can be made for estimands (3) and (4). However, what is remarkable is that despite the fact that these are associational data only, 17Lands data have been widely and effectively used to inform decisions in draft as though they were causal estimates [24–26]. Alex Nikolic, a draft specialist and content creator [27] wrote that GIHWR "is the gold standard for when you're just trying to look up the 'raw power' of a card." [26] Sam Black, former professional player with multiple accolades playing draft [28] wrote that negative IIH "means drawing (this card) actively hurts your chances of winning." [25] While these people are experts at drafting, not observational causal inference scientists, their perspectives do duly consider the issues of treating GIHWR and IIH as causal estimates. As Nikolic claimed that GIHWR is "the least biased metric to determine 'raw power'" while criticizing IIH as "naturally harbors a few biases which makes it an unreliable metric to look at." [26] Black explains cards with negative IIH but high GIHWR as "this card itself isn't as strong as the win rate suggests; it just happens to make it into good decks." [25] Sierkovitz, another draft expert, cautioned that "we can't say from these raw WR numbers whether it is the card in question that may help you win or if it was the rest of the deck doing well," but also recommends to "think of the data as a humble servant that lets you improve your draft and gameplay." [24] Thus, while these sources recognize the weaknesses of GIHWR and IIH as causal estimates, they nonetheless agree that using them to inform picks, especially early ones, is causally informative of a player's chances of winning.

## 4. Sufficiency and necessity of embarrassingly causal qualification for causal observational studies

### 4.1. Sufficiency of embarrassingly causal contexts

The case described is qualified as an "embarrassingly causal" scenario. An embarrassingly causal scenario is one in which the main exposure → outcome relationship of interest is so supported by domain knowledge that the reasonableness of including it in a DAG for observational causal inference can be safely assumed. Naturally, the true sign or magnitude of this assumed causal effect, including the possibility that it is exactly zero, remains an empirical question. Rather, we hold that the embarrassingly causal quality is a sufficient condition for justifying the attempt to estimate the effect in purely observational data settings.

All the data in 17Lands.com are observational only. There is no randomization or control made in its collection. While useful estimands (1) – (4) can be conceptualized, they are very difficult if not impossible to operationalize for direct estimation. Despite this, no reasonable draft player would doubt the causal direction $C_t \to 1_{win}$; whether they choose card $c$ on some pick $t$ causally affects the outcomes of the games in that draft that they have yet to play. This domain knowledge is so strongly accepted that the expected values of GIHWR and IIH are simply assumed to be informative estimands even if those expectations are not (3) or (4) respectively, and such that simply picking the card with the highest GIHWR at $t = 1$ is a valid causally informed decision for maximizing win rate.

This example is not intended to suggest that embarrassingly causal scenarios justify relaxing rigorous methods for establishing and estimating useful causal estimands, but rather illustrates how this quality bestows confidence that even purely associational data, whatever their actual values end up being, are causally informative. As such, we claim that the embarrassingly causal quality is a sufficient condition for justifying the necessary assumption of causal relevance in causal observational studies. That is, for scenarios where the relationship between the proposed exposure and outcome is embarrassingly causal, it is justified to pursue constructing the estimand and its estimator and acquiring the observational data that would allow for estimation.

### 4.2. Extent of necessity of embarrassingly causal contexts

The embarrassingly causal qualification is by no means a low bar for a proposed observational causal inference study to clear. As illustrated by the case examined here, it requires overwhelming domain-level acceptance that, in principle, the exposure → outcome relationship holds. The causal direction itself is not in dispute; the scientific question concerns the magnitude, sign, or contextual relevance of the effect within the data-generating process at hand, including the possibility that the effect is negligible or exactly zero.

Therefore, in considering the necessity of embarrassingly causal qualification, we first provide more examples of situations that fit this context. Consider the very common subset of observational causal inference studies that seek to examine the impact of interventions from successful randomized controlled trials (RCTs) in "real world evidence" (RWE) settings. In such studies, the causal direction from intervention to outcome is generally not plausibly contested unless the internal validity of the RCT itself is challenged. That is, if one is sufficiently satisfied that the RCT showed causality of exposure to outcome for the sample in the study, the fact that this sample is only expected to generalize for the population it specifically represents still implies that it can plausibly transport to other populations it does not exactly represent, thereby justifying data gathering and analysis in those external settings under some suitable observational causal inference strategy. Thus, this subset of observational causal inference studies does fall under the embarrassingly causal qualification. This is consistent with prevailing opinions on the utility of real world evidence studies in supplementing randomized clinical trials [29, 30]. The U.S. FDA provides guidance on the use of RWE to help support approval of a new indication for drugs that have already been approved on the basis of clinical trial evidence [30].

Another subset of scenarios that qualify as embarrassingly causal is when the exposure → outcome relationship has a universally supported mechanism. For example, consider examining the effect of radiation exposure on getting cancer. The mechanism for the basic causal claim – exposure to radiation causally contributes to the risk of getting cancer, is well-known and accepted [31]. This applies across different settings where the same relationship is investigated, like in the INWORKS cohort study that was recently updated with causal estimands for the direct effect of radiation dose on cancer mortality [32]. This same reasoning can be used to justify why studies that investigate asbestos exposure effects on developing mesothelioma [33] or alcohol exposure effects on liver cirrhosis are likewise embarrassingly causal [34].

Having identified broader examples of scenarios that qualify as embarrassingly causal, we now consider how necessary this qualification is for justifying causal interpretation in observational studies. Importantly, the absence of this quality does not hinder the analysis of observational data, nor does it preclude the use of explicitly causal frameworks, including approaches that invoke causal structure such as instrumental variables designed to approximate randomization or structural equation models used to define causally motivated estimands. However, when the embarrassingly causal quality is absent, it means there remains substantive doubt that the assumed causal structure is correct, with alternative causal structures, especially unmeasured confounding or reverse causation, that are considered at least as plausible. This uncertainty impacts the interpretation of any resulting causal estimate regardless of its sign or magnitude, because the estimate is conditional on a causal structure that is itself not uniquely supported by domain knowledge. This is especially concerning for causal discovery and DAG-exploration algorithms, which attempt to infer the causal structure directly from observational data. We argue similarly with other critiques of these methods [35, 36] that in the absence of embarrassingly causal structure, such approaches are unavoidably underdetermined: multiple, substantively distinct causal graphs remain observationally compatible with the data, and any edge orientation relies on strong auxiliary assumptions rather than domain-supported causal direction. As a result, these methods do not resolve causal ambiguity, and estimates resulting from them should not satisfactorily rebut contrary plausible causal explanations that may simply not yet have available data.

5. Conclusions and Recommendations

Through the case of purely observational data used to make causal decisions in Magic: The Gathering drafts, we showed the importance of the causal relevance assumption for the exposure → outcome relationship that underlies any attempt to draw causal inference from observational data. From this, we introduced the criterion of embarrassingly causal scenarios to describe settings in which this assumption is overwhelmingly supported by domain knowledge, including mechanistic understanding or accumulated empirical evidence. We claim that qualification of an exposure → outcome relationship as embarrassingly causal is a sufficient condition for justifying the positing of a corresponding causal structure, including expansion into a DAG with outcome-relevant estimands.

Consequently, these conclusions provide some formal guidance for authors, reviewers, and readers of research studies that infer about causal effects from observational data. For authors, the proposed sufficiency condition provides a principled basis for justifying the inclusion of causal edges which qualify as embarrassingly causal. Magic: The Gathering was presented as an illustrative example, but it is not unique in this regard. Clear evidence of a causal relationship existing from conducted RCT serves as embarrassingly causal justification for including those corresponding causal edges in subsequent observational studies.

For reviewers, this paper provides some guidance on how to evaluate assumptions made by observational studies that claim causal effects. When authors justify inclusion of causal edges based on supposed subject-matter expertise, reviewers should examine if subject matter experts regard the relationship between the proposed exposure and outcome as embarrassingly causal. If this is not the case, then it is prudent for reviewers to have authors accurately represent competing arguments from subject-matter experts on why this relationship is not embarrassingly causal. Where reviewers are themselves, subject-matter experts, this presents a fortuitous situation as the proposed causal edge should ideally be embarrassingly causal from their perspective. If not, then their critique as to why should be provided towards informing the manuscript regardless of how large or statistically significant the effects estimated for those edges are. Naturally, even if a causal edge is considered embarrassingly causal, it still falls upon the authors to provide a valid estimating procedure. Embarrassingly causal only provides justification for including the causal edge and thus proceeding to the next step of properly proposing an estimand and estimating it.

For readers, looking for how well assumptions for key causal edges are framed as embarrassingly causal when reading a research paper can help regard the evidentiary value of everything that follows from that paper about that causal edge. While the embarrassingly causal qualification is not necessary for exploring observational data, its absence should caution readers that any resulting causal estimates are conditional on structural assumptions that are not overwhelmingly supported by domain knowledge and therefore cannot, on their own, rule out competing causal explanations, such as those involving unobserved variables, regardless of the size of effects found.